# Joint Optimization Framework for Operational Cost Minimization in Green Coverage-Constrained Wireless Networks

Ganesh Prasad, Deepak Mishra, and Ashraf Hossain

## Abstract

In this work, we investigate the joint optimization of base station (BS) location, its density, and transmit power allocation to minimize the overall network operational cost required to meet an underlying coverage constraint at each user equipment (UE), which is randomly deployed following the binomial point process (BPP). As this joint optimization problem is nonconvex and combinatorial in nature, we propose a non-trivial solution methodology that effectively decouples it into three individual optimization problems. Firstly, by using the distance distribution of the farthest UE from the BS, we present novel insights on optimal BS location in an optimal sectoring type for a given number of BSs. After that we provide a tight approximation for the optimal transmit power allocation to each BS. Lastly, using the latter two results, the optimal number of BSs that minimize the operational cost is obtained. Also, we have investigated both circular and square field deployments. Numerical results validate the analysis and provide practical insights on optimal BS deployment. We observe that the proposed joint optimization framework, that solves the coverage probability versus operational cost tradeoff, can yield a significant reduction of about 65% in the operational cost as compared to the benchmark fixed allocation scheme.

## Index Terms

Base station location management; coverage probability; network operational cost; power allocation; optimization; green network deployment

G. Prasad and A. Hossain are with the Department of Electronics and Communication Engineering, National Institute of Technology, Silchar, India (e-mail: {gpkeshri, ashraf}@ece.nits.ac.in).

D. Mishra is with the Department of Electrical Engineering, Indian Institute of Technology Delhi, 110016 New Delhi, India (e-mail: deepak.mishra@ee.iitd.ac.in).

# I. Introduction and Background

Today with evolution of various applications based on digital world, the number of UEs and demand of data traffic are increasing exponentially without any compromise with the serving coverage quality of the UEs. Regarding improvement in coverage over the field, various works have been done on deployment strategy of the BSs over the field. The conventional grid model with all the cells being hexagonal in shape and occupying equal area has been shown to be less tractable in a practical environment [2]. Although, deployment of BSs based on homogeneous Poisson point process (HPPP) and BPP is more tractable for satisfying the practical aspects [3], [4], deterministic deployment of BSs according to distribution of the UEs is more realistic and has been shown to have better performance [5].

In the modern world, the data traffic increases almost a factor 10 every 5 years [6]. This causes a huge increment in infrastructure cost every year for meeting the desired Quality of Service (QoS). This increment in infrastructure causes significant increase in power dissipation by $16\%$-$20\%$ per year which results in consumption of $180$ billion kWh electricity per year, which is nearly $1\%$ of the world-wide total energy consumption. These huge consumptions of energy result in carbon dioxide ($CO_2$) and other greenhouse gases emission of nearly 130 million tons every year [6]. This has led to an indispensable need for QoS-constrained green network deployment strategies that maximize utility of operational cost incurred in achieving a desired coverage demand of all the users intended to be served.

## A. Related Works

*Deployment Models:* Currently, most of the literature on deployment of BSs are modeled on the distribution of BSs and UEs by HPPP for practical environment. In [3], the authors show that deployment of BSs and UEs by independent HPPP is more tractable and satisfy the practical aspects than placing the BSs on a grid by conventional methods. A survey on modeling of multi-tier cellular networks and cognitive networks have been done in [2] using stochastic geometry, where according to type of the network and Media Access Control (MAC) layers, various point processes like Poisson point process, BPP, hard core point process, and Poisson cluster process and their performances have been discussed. However in [4], it was argued that binomial point process (BPP) is more realistic and practical network model as compared to HPPP. In contrast to HPPP, in BPP a known and finite number of UEs are distributed in a field. So for better accuracy, we consider a practical setting where UEs are deployed following a BPP.

*Power Allocation:* One of the method for reduction of power consumption is to dynamically turn BSs on/off based on the time and spatial distribution of the traffic load. Various methods for deciding the sleeping mode of the BSs have been discussed in [7]–[15]. Authors in [7] and [8] considered the switching of the BSs based on the traffic profile. Whereas in [9], both the traffic profile and density of the BSs are considered for deciding the switching. Authors in [10] proposed a switching-based energy saving algorithm, which can achieve energy savings of up to $80\%$. However, these works [7]–[10] did not consider any Quality of Service (QoS) constraint to be met while minimizing the energy cost. Authors in [11] discussed about the trade off between energy saving and spectral efficiency due to the switching of BSs, and thereby designed an optimal control mechanism to solve this trade off. In [12], both centralized and decentralized BS energy saving schemes have been proposed under the constraint of outage probability. Authors in [13] and [14] investigated the impact of sleep operation on the blocking probability and delay, respectively. A survey in [15] gives the state of the art on the proposals for reducing the power consumption at the BSs by implementing sleep operations. Although, the works in [7]–[15] optimize the BSs densities by efficiently controlling the switching operation, they have not discussed about the joint optimization of transmit power and location of the BSs.

*BS Localization:* Energy efficient network designs by optimizing the BSs densities without any switching of BSs have been studied in [16]–[21]. In [18], an energy efficient network is designed by optimizing the densities of BSs without any QoS constraint. Whereas in [22], blocking probability has been taken as a constraint in the network design. The coverage probability variation with BS density is studied in [16] for optimizing the operational power cost of the network. In [19], the optimal combination of macro BSs and micro BSs was investigated for satisfying a minimum data traffic demand. On the similar lines, authors in [17] proved that the network power consumption can be reduced by finding the smallest set of BSs that can sustain a given data traffic load. In [20] and [21], the per unit area power consumption of the network is minimized by optimizing the densities of BSs under the constraint of coverage and data rate. However, these works [16]–[21] did not consider the transmit power optimization at BSs while considering the practical deployment constraints.

*Operation Cost Minimization:* There have been some recent developments [23]–[26] for improving the operational power cost by optimizing more than one parameter of the network. In [23], this improvement is achieved by optimizing the BS densities, their transmit power, and deployment factor of the BSs. A method for reducing the power consumption by optimizing the transmit power, BSs densities, number of BS antennas, and number of UEs per cell in a network, has been proposed in [24]. Considering the joint optimization of BSs density and transmit power under the coverage constraint, it was shown in [25] that coverage performance of the system converges to a fixed value with energy related deployment factor. Authors in [26] first optimized the location and power allocation at the BSs, and then separately optimized the count and location of the BSs. Yet, the joint optimization of number of BSs, their transmit power, and location has not been investigated while incorporating BPP model for UEs.

## B. Motivation and Key Contributions

Although most of the works considered the optimization of randomly deployed BS densities, it has been noted that the deterministic deployment of BS is more realistic and has better performance [5]. *To the best of our knowledge, this is the first work that considers coverage-constrained operational cost minimization by jointly optimizing the number of BSs, their transmit power, and locations while considering a realistic BPP for deployment of UEs.* Also, the coverage constraint has been applied to the statistically farthest UE in a cell because each BS takes the responsibility for coverage of all associated UEs. Key contributions of this work are as follows.

- Considering a realistic environment for UEs deployment, we have formulated a coverage constraint joint optimization problem for minimization of the operational power cost. Due to its nonconvex and combinatorial nature, a non-trivial solution methodology is proposed that decouples the joint problem into three individual optimization problems and provides an efficient way to obtain the joint optimal solution.

- We have considered both circular and square field deployments for operational cost minimization while satisfying an average coverage demand. Joint optimal solutions are obtained in each case and the impact of asymptotically high and moderate densities of UEs on the localization of BSs is also discussed. Further, we have discussed the method to derive the distribution of the distance between a BS an its associated UE for differently shaped cells. This distance distribution is used to obtain the coverage probability of the farthest UE from its BS inside a cell and this result is validated via extensive Monte-carlo simulations.

- For minimization of power cost over the network, first we jointly optimize the sectoring type involved in the cells generation and the associated location of the BS inside each cell. Here we have shown that the optimal BS localization is based on the minimization of farthest point Euclidean distance in each cell.

- A tight near-optimal analytical approximation for the optimal power allocation is obtained as a function of the underlying BS deployment. We have shown via numerical investigation that this approximation is very tight under practical system constraints and very tightly matches with the global optimal power allocation for high QoS applications having very high coverage quality demand.

- With both optimal BSs location and transmit power allocation obtained as a function of number $N_B$ of BSs, we prove that the resulting single variable operational cost is unimodal in $N_B$. Using this property an efficient iterative scheme is presented to obtain the optimal number of BSs that in turn yields the optimal BS localization and transmit power allocation providing the minimized operational cost required to meet the underlying coverage demand of each UE in the network.

- Numerical investigation is carried out to validate the analysis and gain nontrivial insights on the impact of various system parameters on the optimized average coverage quality versus cost incurred trade off. A comparison study of operational cost minimization in the square and circular fields having same area is also carried out. Finally, to corroborate the importance of the proposed joint optimization framework, we present its performance comparison against the benchmark schemes to quantify the achievable reduction in operational cost.

## C. Paper Organization

The rest of the paper has been organized as follows. In Section II, we discuss the system model detailing the network topology, wireless channel and cost models. Based on that, we obtain an expression for the average coverage probability of the farthest UE in a cell and then formulate the proposed joint optimization framework in Section III. The optimal BS deployment strategy is outlined in Section IV. Using this, the tight approximation for optimal transmit power allocation of BSs along with the optimal number of BSs to be deployed for meeting a coverage demand are derived in Section V. A method for obtaining the coverage-constrained minimized operational cost in a square field has been discussed in Section VI. Numerical results validating the analysis and providing insights on the joint optimal deployment strategy are presented in Section VII. Lastly, the paper is concluded in Section VIII.

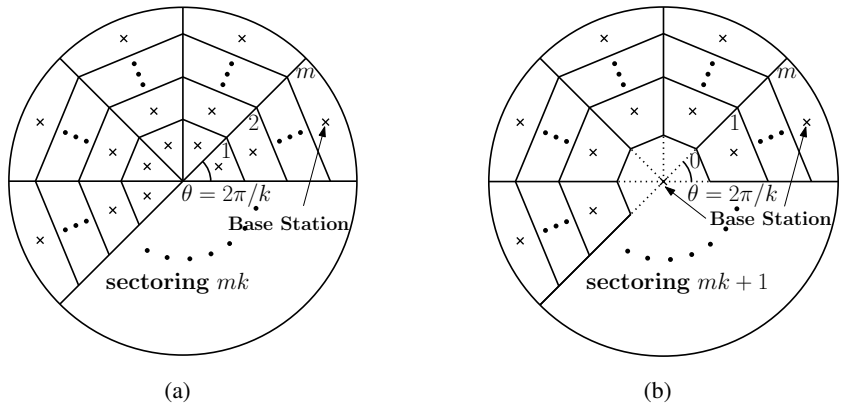

Fig. 1: Generation of cells in a circular field based on the sectoring type (a) $\mathcal{M} = mk$, (b) $\mathcal{M} = mk + 1$. $\theta$ denotes the angle of a sector.

## II. System Model

In this section, we first introduce the considered network topology for deployment of BSs and UEs over the circular field. Next, we present the channel model adopted for downlink communication from BSs to UEs. Lastly, the adopted power cost model of the BSs for characterizing the operational cost is presented.

### A. Network Topology

We consider a homogeneous cellular network deployment, where $N_U$ UEs form a BPP by their independent uniform distribution in a circular field. $N_B$ BSs are deterministically deployed over this field for meeting the required average coverage quality for each UE. Under the assumption of mitigation of interference from intracell and intercell downlink communication, the BSs are assumed to employ the orthogonal multi-access techniques [27]. Each of the BSs and UEs are equipped with single omnidirectional antenna and the downlink association of an UE to a BS in a cell is based on Voronoi-tessellation [28]. Following this, we propose an efficient cell generation method for the circular field to ensure there are no coverage holes and the distance of the farthest point in a cell from its BS is reduced maximally. Here, we first divide the circular field into equal sectors of same angle $\theta$, and then in each sector the BSs are placed along the symmetric line in the radial direction as shown in Fig. 1. Below we define the two sectoring types considered for optimal deployment of BSs over the circular field.

*Definition 1:* The cells in a circular field are generated by considering two sectoring $\mathcal{M} = N_B$ types. (a) $\mathcal{M} = mk$, where the circular field is divided into $k$ equal sectors (each of angle $\theta = 2\pi/k$), and in each sector, $m$ BSs are deployed along its symmetric line. (b) $\mathcal{M} = mk + 1$, where the circular field is divided into $k$ equal sectors (each of angle $\theta = 2\pi/k$) and along with the placement of a BS at the center of the circular field, $m$ BSs are deployed along its symmetric line in each sector.

The major difference between sectoring $\mathcal{M} = mk$ and $\mathcal{M} = mk + 1$ is that in the later case, a BS is deployed at the center. All the sectors generated in sectoring $\mathcal{M}$ have same properties because each of them is generated by dividing the circular field in equal angle $\theta = 2\pi/k$. Therefore, it is sufficient to investigate the optimized performance of any one of them. The BSs are deployed along the radial direction in a sector and their locations from the center of the circular field is given by (a) $\mathbf{d} = \{d_j; j \in \{i, i+1, \ldots, m\}\}$ where $i = 1$ for $\mathcal{M} = mk$ (cf. Fig. 1(a)) and $i = 0$ for $\mathcal{M} = mk + 1$ (cf. Fig. 1(b)), where $d_j$ is the location of the BS from the center of the field in $j^{th}$ cell of a sector.

## B. Channel Model

The received power from the BS is assumed to face path loss and frequency selective Rayleigh fading. Thus, if the distance of $n^{th}$ nearest UE from its associated BS in cell $i$ is $r_{n,i}$, then the corresponding channel power gain is given by $h_{n,i} r_{n,i}^{-\alpha}$, where $h_{n,i}$ is random channel fading gain following the exponential distribution with unit mean and $\alpha$ is the path loss exponent. The signal-to-noise ratio (SNR) received at that UE is given by $\gamma_{n,i} = \frac{P_t h_{n,i}}{\sigma^2 r_{n,i}^{\alpha}}$, where $P_t$ is the transmit power of each BS and $\sigma^2$ is the variance of the zero mean additive white Gaussian noise signal received at each UE. The coverage of the $n^{th}$ nearest UE from the BS depends on whether the received SNR $\gamma_{n,i}$ at that UE is greater than the threshold $T$ required for successfully detecting the information content in the received signal. The coverage probability of the $n^{th}$ nearest UE from the BS is $\mathbf{Pr}\left[\gamma_{n,i} \geq T\right]$. Using it along with the definition of $\gamma_{n,i}$, corresponding average coverage probability is given by:

$$P_{\text{cov}}^{n,i} = \mathbb{E}_{r_{n,i}} \left\{ \mathbf{Pr}\left[ \frac{P_t h_{n,i} r_{n,i}^{-\alpha}}{\sigma^2} \geq T \right] \right\} = \int_0^{r_{u,i}} e^{-\frac{T\sigma^2 r_{n,i}^{\alpha}}{P_t}} f_{n,i}(r_{n,i}, \mathbf{d_i}) \, dr_{n,i}, \qquad (1)$$

where $\mathbb{E}_{r_{n,i}}\{\cdot\}$ represents the expectation with respect to the random distance $r_{n,i}$ of $n^{th}$ nearest UE, $f_{n,i}(r_{n,i}, \mathbf{d_i})$ is the probability density function (PDF) of $r_{n,i}$ with $r_{u,i}$ as the upper limit for $r_{n,i}$, and $\mathbf{d_i} = \{d_{i-1}, d_i, d_{i+1}\}$ is the location of the BSs in $(i-1)^{th}$, $i^{th}$, and $(i+1)^{th}$ cells respectively of a given sector of angle $2\pi/k$. We notice that the average coverage probability of the $n^{th}$ nearest UE from the BS in $i^{th}$ cell depends not only on the location of its own BS, but also on the location of the BSs in adjacent $(i-1)^{th}$ and $(i+1)^{th}$ cells, because the boundaries of the cells are determined by the Voronoi diagram. However, in case of $1^{st}$ and $m^{th}$ (last) cells, the boundaries along the sector depends only on location of the BSs in $1^{st}$, $2^{nd}$ and $(m-1)^{th}$, $m^{th}$ cells, respectively because, one of their boundaries is fixed along the sector. Now if the average coverage probability of the farthest UE from the BS in a cell satisfies a given coverage demand, then statistically it will also be satisfied by the other UEs inside the cell. Hence, in the proposed analysis, we have applied the average coverage constraint only on the farthest UE in a cell.

As our main goal is to minimize the operational cost of the BSs required to meet an average coverage demand, we next present the power cost model for the BS deployment.

## C. Operational Cost Model for the BS Deployment

From [25], [29], the power consumption model for a BS to analyze the power dissipation in downlink transmission is given as:

$$P_{BS} = N_{PA} \left( \frac{P_t}{\mu_{PA}} + P_{SP} \right) (1 + C_{PCB}),$$ (2)

where $P_{BS}$ is the total power consumed by a BS which constitutes of (i) transmit power $P_t$ by the BSs, (ii) power amplifier (PA) efficiency $\mu_{PA}$, (iii) total number of power amplifiers $N_{PA}$, (iv) power dissipation $P_{SP}$ in signal processing on the data, and (v) cost $C_{PCB}$ due to power supply, cooling, battery backup and other maintenance costs. This can be further simplified as linear cost model for a BS as:

$$P_{BS} = a_B P_t + b_B,$$ (3)

where $a_B$ representing the coefficient of power consumption that scales the radiated power and $b_B$ accounting for other consumptions due to signal processing, cooling, and battery backup. Thus, the total operational power cost is $N_B P_{BS}$ [25] and it has to be minimized for enabling coverage-constrained green communications.

## III. Problem Formulation

In this section we first obtain average coverage probability of farthest UE as a function of location of the BSs for a given $N_B$, $\mathbf{d_i}$, $\mathcal{M}$, and the field dimensions. Later, we present the mathematical formulation for the joint optimization problem to minimize the operational cost.

## A. Average Coverage Probability of the Farthest UE

As discussed in Section II-A, we apply the average coverage constraint on the farthest UE inside a cell and it depends on the distribution of the distance $r_{\text{far},i}$ of the farthest UE. Its cumulative density function (CDF) $F_{\text{far},i}(r_{\text{far},i}, \mathbf{d_i})$ and PDF $f_{\text{far},i}(r_{\text{far},i}, \mathbf{d_i})$ are given by:

$$F_{\text{far},i}(r_{\text{far},i}, \mathbf{d_i}) = \sum_{k=0}^{N_U} \binom{N_U}{k} \left(\frac{A_i}{W}\right)^k \left(1 - \frac{A_i}{W}\right)^{(N_U-k)} [F_i(r_{\text{far},i}, \mathbf{d_i})]^k, \tag{4a}$$

$$f_{\text{far},i}(r_{\text{far},i}, \mathbf{d_i}) = \sum_{k=0}^{N_U} \binom{N_U}{k} \left(\frac{A_i}{W}\right)^k \left(1 - \frac{A_i}{W}\right)^{(N_U-k)} k f_i(r_{\text{far},i}, \mathbf{d_i}) [F_i(r_{\text{far},i}, \mathbf{d_i})]^{(k-1)}, \tag{4b}$$

where $F_i(r_{\text{far},i}, \mathbf{d_i})$ is the CDF of $r_{\text{far},i}$ for an UE associated with $i^{th}$ cell, $f_i(r_{\text{far},i}, \mathbf{d_i})$ is the corresponding PDF, $A_i$ is the area of the $i^{th}$ cell, and $W$ is the whole area of the circular field [4]. If the shape of a cell is polygon, then $F_i(r_{\text{far},i}, \mathbf{d_i})$ and $f_i(r_{\text{far},i}, \mathbf{d_i})$ can be obtained by the method discussed in [30] and if the boundary of the cell has circular arc, then it can be calculated using appendix A. So, using (1) and (4), the average coverage probability of farthest UE is given by

$$P_{\text{cov}}^{\text{far},i} = \int_0^{r_{u,i}} e^{-\frac{T\sigma^2 r_{\text{far},i}^\alpha}{P_t}} f_{\text{far},i}(r_{\text{far},i}, \mathbf{d_i}) \, dr_{\text{far},i}. \tag{5}$$

From (5), it is evident that average coverage probability of the farthest UE in $i^{th}$ cell not only depends on the location of its own BS, but also depends on the locations of the BSs in its neighboring $(i-1)^{th}$ and $(i+1)^{th}$ cells in the sector.

## B. Optimization Formulation

Now we present the joint optimization framework for finding the optimal number $N_B$ of BSs to be deployed along with their transmit power allocation $P_t$ and locations $\mathbf{d}$ inside a sector, that minimize the operational cost incurred in meeting an average coverage constraint at the farthest UE in each cell.

$$
\begin{aligned}
\text{(P0)}: &\underset{N_B, P_t, \mathbf{d}}{\text{minimize}} \ N_B \left[a_B P_t + b_B\right] \\
&\text{subject to} \ \ C1: P_{\text{cov}}^{\text{far},i} \geq 1 - \epsilon, \ \forall i \in \mathcal{I}, \\
&\qquad\qquad\ \ C2: N_B = \{1, 2, \ldots, N_{B,\text{max}}\}, \\
&\qquad\qquad\ \ C3: 0 \leq P_t \leq P_{t,\text{max}}, \\
&\qquad\qquad\ \ C4: 0 \leq d_i \leq d_{i,\text{max}}, \ \forall i \in \mathcal{I},
\end{aligned}
\tag{6}
$$

where $\mathcal{I} = \{j, j+1, \ldots, m\}$ in which $j=1$ for $\mathcal{M} = mk$ and $j=0$ for $\mathcal{M} = mk+1$. The constraint $C1$ ensures that the average coverage probability of the farthest UE is greater than or equal to an acceptable threshold $1 - \epsilon$ in each cell. Here $0 < \epsilon \ll 1$ is decided based on the acceptable threshold that enables a minimum required coverage probability. The convex linear constraints $C3$ and $C4$ represent the boundary conditions for $P_t$ and $d_i$, respectively. Here, $P_{t,\max}$ and $d_{i,\max}$ respectively represent the upper bounds on $P_t$ and and location $d_i$ of BS in $i^{th}$ cell. In general, as (P0) is a nonconvex combinatorial optimization problem due to the presence of integer variable $N_B$ and non-convexity of the objective function and constraints $C1$ and $C2$, it is hard to solve it in its current form. So, we present a nontrivial solution methodology that effectively solves this problem by decoupling it into three individual optimization problems. We next obtain the optimal BS deployment strategy for a given number $N_B$ of BSs.

## IV. Optimal Deployment Strategy of BSs

Here we present deployment strategy of BSs over the circular field for a given number $N_B$ of BSs. First, we discuss the optimal BSs deployment when number of UEs in the field is asymptotically very high. After that we carry forward the discussion for scenarios with moderate UEs density. Lastly, we demonstrate that optimal deployment strategy of BSs and selection of sectoring type are based on minimizing the farthest point Euclidean distance in each cell.

### A. Asyptotically High Density of UEs

When the number of UEs $N_U$ over a finite circular field is asymptotically very high ($N_U \to \infty$), then it can be easily shown that in any sub-field of the field, there will be infinite number of UEs, i.e., if $\chi_i$ number of UEs is lying in $i^{th}$ cell (sub-field) of the circular field, then $\chi_i \to \infty$ for $N_U \to \infty$. Therefore, the CDF $F_{\mathrm{far},i}(r_{\mathrm{far},i}, \mathbf{d_i})$ of distance of farthest UE from location $d_i$ of the BS in $i^{th}$ cell can be expressed as: $F_{\mathrm{far},i}(r_{\mathrm{far},i}, \mathbf{d_i}) = \lim_{\chi_i \to \infty}[F_i(r_{\mathrm{far},i}, \mathbf{d_i})]^{\chi_i}$. As CDF $F_i(r_{\mathrm{far},i}, \mathbf{d_i}) = 1$ for $r_{\mathrm{far},i} = r_{u,i}$ and $< 1$ for $r_{\mathrm{far},i} = r_{u,i} - \nu$ ($\nu > 0$), the corresponding probability of lying of farthest UE over the range $r_{\mathrm{far},i} \in (r_{u,i} - \nu, r_{u,i}]$ is given as:

$$\mathbf{Pr}(r_{u,i} - \nu < r_{\mathrm{far},i} \leq r_{u,i}) = \lim_{\chi_i \to \infty}[F_i(r_{u,i}, \mathbf{d_i})]^{\chi_i} - \lim_{\chi_i \to \infty}[F_i(r_{u,i} - \nu, \mathbf{d_i})]^{\chi_i}$$
$$= 1 - 0, \tag{7}$$

where $\nu$ is a very small positive constant ($\nu \to 0^+$) and $r_{u,i}$ is the farthest point Euclidean distance from the BS. Using (7), the PDF of the farthest UE can be written as $f_{\mathrm{far},i}(r_{\mathrm{far},i}, d_i) = \delta(r_{\mathrm{far},i} - r_{u,i})$ for $\chi \to \infty$, where $\delta(\cdot)$ is a Dirac delta function. Therefore, if number of UEs in a field is very high, then farthest UE lies at the farthest point Euclidean distance from the BS. So, we determine the optimal location of the BS by minimizing the farthest point Euclidean distance $r_{u,i}$ from it.

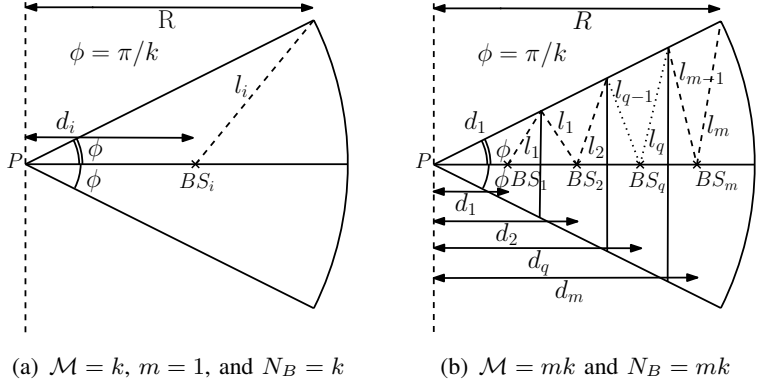

(a) $\mathcal{M} = k$, $m = 1$, and $N_B = k$

(b) $\mathcal{M} = mk$ and $N_B = mk$

Fig. 2: Deployment of BSs over the circular field with two sectoring types.

The conceptual insight for optimal deployment strategy based on minimization of $r_{u,i}$ can be better understood by taking an example of sectoring $k$ with $m = 1$, where one BS deployed in each sector as shown in Fig. 2(a). Here, the farthest point from the BS in cell $i$ (here $i = 1$) is decided among the distances from the three vertices. As out of them, two vertices are at same distance from the BS due to symmetry, the minimum value of $r_{u,i}$ can be obtained by minimizing the maximum of the two distances with respect to $d_i$, i.e., $\min r_{u,i} = \min_{d_i} \max\{d_i, l_i\}$. For $N_B \geq 4$, the minimum value of $r_{u,i}$ is obtained by $l_i = d_i$ which gives $\min r_{u,i} = l_i = d_i = \frac{R}{2\cos\left(\frac{\pi}{N_B}\right)}$ and corresponding optimal location $d_i^* = \frac{R}{2\cos\left(\frac{\pi}{N_B}\right)}$, where $l_i = \sqrt{d_i^2 + R^2 - 2d_i R \cos\phi}$ for a circular field with radius $R$ and $\phi = \theta/2 = \pi/N_B$, for $k = N_B$. For $N_B = \{1, 2\}$, it can be easily shown that $\min r_{u,i} = R$ and $d_i^* = 0$, whereas for $N_B = 3$, $\min r_{u,i} = R\sin\phi$ and $d_i^* = R\cos\phi$ obtained by $\frac{\partial l_i^2}{\partial d_i} = 0$.

Generalizing this concept, if the BSs are deployed by sectoring $\mathcal{M} = mk$ as shown in Fig. 2(b), then optimal location of $m$ BSs in each sector is determined by solving $\min_{\mathbf{d}} \max\{d_1, l_1, l_2, \ldots, l_m\}$, which after similar simplification like for $m = 1$, reduces to $d_1 = l_i, \forall i \in \mathcal{I}$. On using the trigonometric relationship of $l_1, l_2, \ldots, l_m$ with $d_1, d_2, \ldots, d_m$, respectively, we can find the minimum value of farthest point Euclidean distance $r_{u,i}$ and corresponding optimal locations $\mathbf{d}^* = \{d_1^*, d_2^*, \ldots, d_m^*\}$ for all BSs. Likewise, we can also find them in sectoring $mk+1$. In Table I, we have listed the minimum value of farthest point Euclidean distance $r_{u,i}$ and corresponding optimal locations $\mathbf{d}^*$ in second and third column respectively for different sectoring types (upto sectoring $3k$). The computation of optimal sectoring type $\mathcal{M}^*$ is based on obtaining the global minimum value of farthest point Euclidean distance $r_{u,i}$ for a given number of BSs $N_B$. In Fig 3, we have depicted the farthest point Euclidean distance $r_{u,i}$ for different sectoring types (upto sectoring $3k$) using the expressions in second column of Table I. It is perceptible that over a certain range of $N_B$, one sectoring type has global minimum value of $r_{u,i}$. For example, if we compare the expression of $r_{u,i}$ in sectoring $k$, $k+1$, and $2k$, $\frac{R}{4\cos^2\left(\frac{\pi}{N_B-1}\right)-1} < \frac{R}{2\cos\left(\frac{\pi}{N_B}\right)}$ for $N_B \geq 4$ and $\frac{R}{4\cos^2\left(\frac{\pi}{N_B-1}\right)-1} < \frac{R}{4\cos\left(\frac{2\pi}{N_B}\right)\cos\left(\frac{4\pi}{N_B}\right)}$ for $N_B \leq 17$ and vice versa. Therefore sectoring $k+1$ has global minimum value for $4 \leq N_B \leq 17$. Similarly, sectoring $2k$ has global minimum value for $18 \leq N_B \leq 19$, but it cannot take the odd integer value, i.e., $N_B = 19$. So, we have included it in the range for sectoring $k+1$, where $r_{u,i}$ has smaller value than in the range for sectoring $2k+1$. Likewise, the range of $N_B$ for other sectoring type has been given in fourth column of the table. Although, Fig 3 is depicted for a circular field with radius $R = 500$ m, it is effectual for any value of $R$ as it only scales the value of $r_{u,i}$ without affecting the range of $N_B$ for different sectoring types. Note that in sectoring $mk$, $\phi = m\pi/N_B$ whereas for sectoring $mk+1$, $\phi = m\pi/(N_B-1)$. In Table I, we have listed the contents upto sectoring $3k$ because the upper bound of number of BSs $N_{B,\max} = 35$ for our performed experiments.

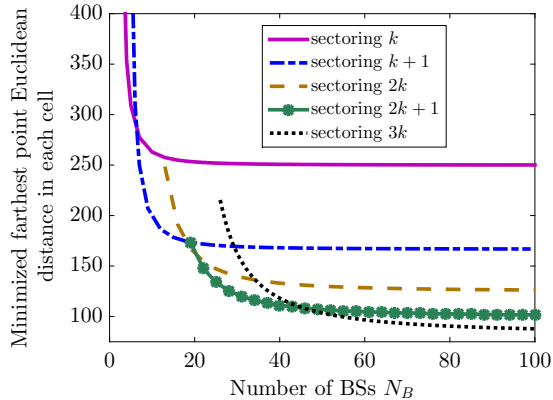

Fig. 3: Variation of minimized farthest point Euclidean distance with $N_B$ in each cell for different type of sectoring.

### B. When the Density of UEs is Moderate

Now, we carry out the optimal deployment strategy in a scenario, when number of UEs $N_U$ in the circular field is moderate (finite). Again, if we give an insight on distribution of distance of farthest UE from the BS in $i^{th}$ cell in which the PDF $f_{\text{far},i}(r_{\text{far},i}, \mathbf{d_i})$ is non-zero for $r_{\text{far},i} \in [0, r_{u,i}]$, i.e., the farthest UE distance depends on farthest point Euclidean distance $r_{u,i}$ from the BS. However, if $r_{u,i}$ gets changed by $\xi > 0$ due to a shift in the location of the BS, then there is a non-zero probability for an UE to lie in the distance range $r_{u,i}$ to $r_{u,i} + \xi$ from the BS. But, for a given value of $N_U$, the farthest UE distance not only depends on $r_{u,i}$, but also on the distribution of area around the boundaries. Therefore, the obtained optimal location $d_i^*$ based on minimization of farthest point Euclidean distance $r_{u,i}$ is different from actual optimal location $d_{i,\text{act}}^*$ which depends on $r_{u,i}$ as well as the distribution of area around the boundaries. For better discernment, we take a peculiar instance, when a single BS is deployed over the circular field as shown in Fig. 4(a). Here, first we find the optimal location $d_0^*$ based on minimization of farthest point Euclidean distance, then a delimited realm around the optimal location $d_0^*$ associated with actual optimal location $d_{0,\text{act}}^*$ of the BS is estimated from which the farthest UE distance is minimum for a given value of $N_U$, where suffix 0 denotes a single BS is deployed over the circular field. The realm associated with $d_{0,\text{act}}^*$ converges to the optimal location $d_0^*$ with increment in $N_U$ and coincides with it, when $N_U \to \infty$.

TABLE I: Optimal sectoring $\mathcal{M}^*$ and corresponding minimized farthest point Euclidean distance and optimal location $\mathbf{d}^*$ of BSs for a given value of $N_B$.

| Optimal sectoring $\mathcal{M}^*$ | Minimized farthest point Euclidean distance | Optimum location $\mathbf{d}^*$ of the BSs | Range of $N_B$ |
|---|---|---|---|
| $k$ | $R\sin\left(\frac{\pi}{N_B}\right)$ | $d_1^* = R\cos\left(\frac{\pi}{N_B}\right)$ | $N_B = 3$ |
| | $\frac{R}{2\cos\left(\frac{\pi}{N_B}\right)}$ | $d_1^* = \frac{R}{2\cos\left(\frac{\pi}{N_B}\right)}$ | $N_B \in \{4,5,6\}$ |
| $k+1$ | $\frac{R}{4\cos^2\left(\frac{\pi}{N_B-1}\right)-1}$ | $d_0^* = 0, d_1^* = \frac{2R\cos\left(\frac{\pi}{N_B-1}\right)}{4\cos^2\left(\frac{\pi}{N_B-1}\right)-1}$ | $N_B \in \{7,8,\ldots,17\}$ $\bigcup\{19\}$ |
| $2k$ | $\frac{R}{4\cos\left(\frac{2\pi}{N_B}\right)\cos\left(\frac{4\pi}{N_B}\right)}$ | $d_1^* = \frac{R}{4\cos\left(\frac{2\pi}{N_B}\right)\cos\left(\frac{4\pi}{N_B}\right)}$, $d_2^* = \frac{R\left(1+\cos\left(\frac{4\pi}{N_B}\right)\right)}{4\cos\left(\frac{2\pi}{N_B}\right)\cos\left(\frac{4\pi}{N_B}\right)}$ | $N_B \in \{18,20,\ldots,44\}$ |
| $2k+1$ | $\frac{R\left(1+2\cos\left(\frac{4\pi}{N_B-1}\right)\right)}{16\cos^2\left(\frac{2\pi}{N_B-1}\right)\cos^2\left(\frac{4\pi}{N_B-1}\right)-1}$ | $d_0^* = 0, d_1^* = \frac{2R\left(1+2\cos\left(\frac{4\pi}{N_B-1}\right)\right)\cos\left(\frac{2\pi}{N_B-1}\right)}{16\cos^2\left(\frac{2\pi}{N_B-1}\right)\cos^2\left(\frac{4\pi}{N_B-1}\right)-1}$, $d_2^* = \frac{4R\left(1+2\cos\left(\frac{4\pi}{N_B-1}\right)\right)\cos\left(\frac{4\pi}{N_B-1}\right)\cos\left(\frac{2\pi}{N_B-1}\right)}{16\cos^2\left(\frac{2\pi}{N_B-1}\right)\cos^2\left(\frac{4\pi}{N_B-1}\right)-1}$ | $N_B \in \{21,23,\ldots,45\}$ |
| $3k$ | $\frac{R\cos\left(\frac{3\pi}{N_B}\right)}{\left(2\cos\left(\frac{6\pi}{N_B}\right)+1\right)\left(\cos\left(\frac{12\pi}{N_B}\right)+\cos\left(\frac{6\pi}{N_B}\right)\right)}$ | $d_1^* = \frac{R\cos\left(\frac{3\pi}{N_B}\right)}{\left(2\cos\left(\frac{6\pi}{N_B}\right)+1\right)\left(\cos\left(\frac{12\pi}{N_B}\right)+\cos\left(\frac{6\pi}{N_B}\right)\right)}$, $d_2^* = \frac{R\cos\left(\frac{3\pi}{N_B}\right)\left(1+2\cos\left(\frac{6\pi}{N_B}\right)\right)}{\left(2\cos\left(\frac{6\pi}{N_B}\right)+1\right)\left(\cos\left(\frac{12\pi}{N_B}\right)+\cos\left(\frac{6\pi}{N_B}\right)\right)}$, $d_3^* = \frac{R\cos\left(\frac{3\pi}{N_B}\right)\left(1+2\cos\left(\frac{6\pi}{N_B}\right)+2\cos\left(\frac{6\pi}{N_B}\right)\right)}{\left(2\cos\left(\frac{6\pi}{N_R}\right)+1\right)\left(\cos\left(\frac{12\pi}{N_B}\right)+\cos\left(\frac{6\pi}{N_R}\right)\right)}$ | $N_B \in \{48,51,\ldots\}$ |

Now, we will estimate the realm of actual optimal location $d_{0,\text{act}}^*$ around the optimal location $d_0^*$ based on minimization of farthest point Euclidean distance from the BS. In a circular field, the farthest point Euclidean distance is minimum when the location $d_0^*$ of the BS is at the center. The delimited realm around optimal location $d_0^*$ associated with its actual optimal location $d_{i,\text{act}}^*$ based on minimization of farthest UE distance from the BS can be estimated by giving an insight on distribution of area around the boundaries, where the farthest UE is lying with a probability $1 - \psi$, where $\psi \in (0, 1)$ is the acceptable threshold. If probability of lying of farthest UE in the range $r_{far} \in [r_u(1-q), r_u]$ from $d_0^*$ is $\geq 1 - \psi$, then it can be expressed as $[F(r_u, \mathbf{d_0^*})]^{N_U} - [F(r_u(1-q), \mathbf{d_0^*})]^{N_U} \geq 1 - \psi$, where $F(r_{\text{far}}, \mathbf{d_0^*})$, $[F(r_{\text{far}}, \mathbf{d_0^*})]^{N_U}$ are the CDF of distance $r_{\text{far}}$ of an UE, farthest UE, respectively, $\mathbf{d_0^*} = \{d_0^*\}$, and $r_u$ is farthest point Euclidean distance from the BS. As CDF $F(r_{\text{far}}, \mathbf{d_0^*}) = \frac{r_{\text{far}}^2}{r_u^2}$ and $F(r_u, \mathbf{d_0^*}) = 1$, after substitution in the expression, we get $q \geq 1 - \psi^{\frac{1}{2N_U}}$. For $N_U \to \infty$, $q \to 0$, i.e., farthest UE distance is same as farthest point Euclidean distance $r_u$ and $d_{0,\text{act}}^*$ coincides with $d_0^*$.

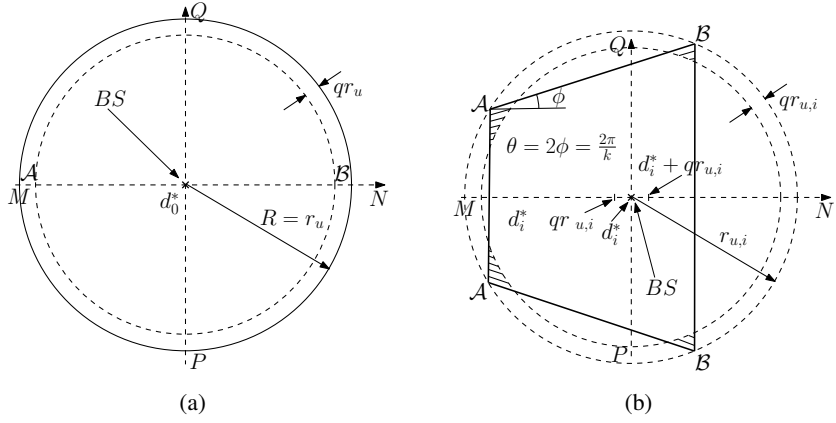

Fig. 4: Actual optimal location of a BS (a) in a circular field, (b) in an $i^{th}$ cell for a given value of $N_U$.

Now, we find the realm around the optimal location $d_0^*$ associated with actual optimal location $d_{0,\text{act}}^*$ of the BS. As the realm of actual optimal location depends on the distribution of area over the width $qr_u$ near to boundaries, i.e., higher area implies higher probability of lying of the farthest UE and vice versa. Therefore, the realm of actual optimal location is nearer to higher distribution of area near to boundaries for minimization of farthest UE distance. As the fields taken in our experiments are $2D$, we take two mutually orthogonal axis MN and PQ as the coordinate system for probing the distribution of area around the boundaries and corresponding realm of actual optimal location from $d_0^*$ which is taken as the origin as shown in Fig. 4(a). Actual location on the axis MN is determined by splitting the field about the orthogonal axis PQ and then we compare the areas $\mathcal{A}$ and $\mathcal{B}$ over the width $qr_u$ on two sides. If $\mathcal{A} > \mathcal{B}$, then probability of lying of farthest UE on left side PQ is more than right side. Therefore, for minimizing the farthest UE distance, the shift will be on left side from $d_0^*$ along the axis MN. Similarly, for $\mathcal{A} < \mathcal{B}$, the shift will be on right side. Due to symmetry of the circular field about the center located at $d_0^*$, $\mathcal{A} = \mathcal{B}$ and so, there is equal probability of lying of farthest UE on two sides over the width $qr_u$ which results in no shift along the axis MN. Similarly the shift can be examined along axis PQ after splitting the field about axis MN which gives zero shift again due to symmetry property of the circular field in all directions. Therefore, the actual optimal location $d_{0,\text{act}}^*$ of the BS coincides with $d_0^*$ for any value of $N_U$.

In general, if we consider an $i^{th}$ cell in the scenario, when multiple BSs are deployed as shown in Fig. 4(b), the value of $q$ can be obtained numerically by $\sum_0^{N_U} \binom{N_U}{k} (\frac{A_i}{W})^k (1 - \frac{A_i}{W})^{N_U - k} (F_i((1 - q)r_{u,i}, \mathbf{d_i^*}))^k \leq \psi$ which is obtained using (4), where $F_i(r_{far,i}, \mathbf{d_i^*})$ is the CDF of distance $r_{far,i}$ of an UE from $d_i^*$ and $d_i^*$ is obtained by minimizing the farthest point Euclidean distance from the BS in the $i^{th}$ cell. As the field is symmetric about the axis MN and asymmetric about the orthogonal axis PQ, the shift will be along the axis MN from the location $d_i^*$ due to the difference in areas $2\mathcal{A}$ and $2\mathcal{B}$ on two sides, where $\mathcal{A}$ and $\mathcal{B}$ are the areas of the field over the width $qr_{u,i}$ near to boundaries as shown in Fig. 4(b). If $2\mathcal{A} > 2\mathcal{B}$, then the realm is on left side over the range $d_{i,\text{act}}^* \in (d_i^* - qr_{u,i}, d_i^*)$, for $2\mathcal{A} > 2\mathcal{B}$, it is on right side over the range $d_{i,\text{act}}^* \in (d_i^*, d_i^* + qr_{u,i})$ and no shift for $2\mathcal{A} = 2\mathcal{B}$, i.e., $d_{i,\text{act}}^* = d_i^*$, where $d_{i,\text{act}}^*$ is actual optimal location over the delimited realm for a given value of $N_U$ in $i^{th}$ cell. Now, we will do a brief discussion on dependence of $d_{i,\text{act}}^*$ on $N_U$ and number of angular sectors $k$. As discoursed before, with increment in $N_U$, $d_{i,\text{act}}^*$ converges to $d_i^*$ and asymptotically coincides with it, when $N_U \to \infty$. The angle of a sector is $\theta = 2\phi = \frac{2\pi}{k}$, therefore with increment in $k$, $\phi$ decreases which results in decrement in difference between heights of the rhombus as shown in Figs. 2(b) and 4(b). Therefore, the difference between the areas $2\mathcal{A}$ and $2\mathcal{B}$ decreases and ideally both are equal, when $\phi = 0$. So, with increment in $k$, $d_{i,\text{act}}^*$ converges to $d_i^*$ due to equal area over the width $qr_{u,i}$ on left and right side. Via extensive simulations, we observe that there is diminishable root mean square error (RMSE), when the BSs are deployed based on minimization of farthest point Euclidean distance $d_i^*$ instead of actual optimal location $d_{i,\text{act}}^*$ based on minimization of farthest UE distance which has been discoursed in Section VII at an instance scenario. Therefore, in the following optimization technique, we will take optimal location of BSs based on minimization of farthest point Euclidean distance for simplicity in calculation.

As the optimal location of BSs in a cell is based on minimum value of farthest point Euclidean distance, the selection of optimal sectoring $\mathcal{M}^*$ is also based on minimum value of farthest point Euclidean distance in each cell for a given value of $N_B$ as discoursed before in Section IV-A.

## V. Proposed Solution Methodology for Operational Cost Minimization

Continuing with our solution methodology of solving (P0) by decoupling it into three individual optimization problems, in this section we find the optimal transmit power $P_t^*$ of BSs and their optimal count $N_B^*$. With optimal location obtained for a given number $N_B$ of BSs in Section IV, now we propose a tight approximation for optimal $P_t$ as a function of optimal BS location $\mathbf{d}^*$ and optimal sectoring $\mathcal{M}^*$ for a given number $N_B$ of BSs. Lastly, using these developments, we discuss the reduction of (P0) to a unimodal single variable optimization problem in $N_B$ that can be solved efficiently to obtain optimal number $N_B^*$ of BSs, which will eventually give optimal localization $(\mathbf{d}^*, \mathcal{M}^*)$ and transmit power $P_t^*$ for $N_B^*$ BSs.

### A. Tight Approximation for Optimal Power Allocation

High QoS applications require very high average coverage probability, i.e., threshold $\epsilon$ is generally very low in practice. Considering this requirement, from constraint $C1$, we note that to ensure an average coverage of atleast $90\%$ (i.e., $\epsilon \leq 0.1$) for any distribution $f_{\text{far},i}(r_{\text{far},i}, \mathbf{d_i})$ of farthest UE's distance from its BS in $i^{th}$ cell, the argument $\frac{T\sigma^2 r_{\text{far},i}^\alpha}{P_t}$ of the exponential term should be $\leq 0.1$. As, $\mathrm{e}^{-x} \approx 1 - x, \forall, x \leq 0.1$ with a percentage error $\leq 0.053\%$ and this approximation error decreases at an exponential rate with decreasing $x$. Applying this approximation to the average coverage probability, constraint $C1$ in (P0) can be rewritten as

$$P_{\text{cov}}^{\text{far},i} \approx 1 - \frac{T\sigma^2}{P_t} \int_0^{r_{u,i}} r_{\text{far},i}^\alpha f_{\text{far},i}(r_{\text{far},i}, \mathbf{d_i}) \mathrm{d}r_{\text{far},i} \geq 1 - \epsilon, \forall i \in \mathcal{I}. \tag{8}$$

Here also, the approximation error is $< 0.053\%$ and it reduces exponentially with decrease in acceptable threshold $\epsilon$; it reduces to zero, when the coverage requirement is $100\%$, i.e., $\epsilon = 0$.

We employ this exponential approximation to obtain a tight approximation for optimal power allocation $P_t^*$ at each BS. Below, $C1$ is rewritten as an approximate lower bound for $P_t$ at optimal location $d_i^*$.

$$P_t \geq \frac{T\sigma^2}{\epsilon} \int_0^{r_{u,i}} r_{\text{far},i}^\alpha f_{\text{far},i}(r_{\text{far},i}, \mathbf{d_i^*}) \mathrm{d}r_{\text{far},i}, \ \forall i \in \mathcal{I}. \tag{9}$$

where $\mathbf{d_i^*} = \{d_{i-1}^*, d_i^*, d_{i+1}^*\}$ are the optimal location of BSs in $(i-1)^{th}$, $i^{th}$, and $(i+1)^{th}$ cells, respectively. Since, the operational cost to be minimized is directly proportional to the transmit power $P_t$, we need to allocate just sufficient transmit power that can help in achieving the desired coverage threshold $1 - \epsilon$. With homogeneous network consideration, the power allocation for all BSs is same and it is obtained by taking the maximum of the power allocations as obtained by solving (9) at strict equality for each BS. Hence, the tight approximation of optimal power allocation is given by

$$P_t^* \approxeq \max_i \left\{ \frac{T\sigma^2}{\epsilon} \int_0^{r_{u,i}} r_{\text{far},i}^{\alpha} f_{\text{far},i}(r_{\text{far},i}, \mathbf{d}_i^*) \mathrm{d}r_{\text{far},i} \right\}, \tag{10}$$

which is a function of number of BSs, their locations, and acceptable threshold $\epsilon$. With transmit power for each BS set as $P_t^*$ defined in (10), constraint $C1$ is implicitly satisfied and the value of optimal $P_t^*$ can be obtained by optimizing the locations $\mathbf{d}^*$ of the BSs and corresponding sectoring type $\mathcal{M}^*$ for a given value of $N_B$.

## B. Efficeint Iterative Scheme to Find Optimal Number of BSs

From the developments in Sections IV and V-A, we note that both optimal BS location along with the corresponding optimal sectoring type and transmit power allocation can be represented as a function of $N_B$. This reduces the multi-variable constrained joint optimization problem (P0) to a univariate problem in $N_B$, where $N_B$ is a positive integer variable to be optimized. Next, we show that this reduced problem possesses the global optimality property in $N_B$.

As with increasing $N_B$ the area of the cell to be covered under a BS approximately reduces by a factor of $\frac{1}{N_B}$, the resulting distance of the farthest UE from the BS and hence, the transmit power $P_t$ required to meet the underlying coverage constraint also decrease by a factor of $\frac{1}{N_B^\beta}$ where $\beta > 0$. Further, the objective function of (P0) is a product of $N_B$ and a affine transformation $a_B P_t + b_B$ of $P_t$. So on relaxing the integer constraint on $N_B$, we note that $N_B$ is a positive linear function and $P_t$ for a given $N_B$ with optimized BS location, as discussed above, is a nonlinear decreasing convex function of $N_B$ because the rate of decrease in $P_t$ is decreasing with increased $N_B$. Using these results in [31, Table 5.2 and Prop. 3.8], the unimodality of the objective in $N_B$ is proved. The method for determining the optimal number of BSs $N_B^*$ and corresponding minimum value of the operational cost is outlined in Algorithms 1 and 2 which are discussed next.

Algorithm 1 outlines a procedure to obtain operational cost as a function $f(N_B)$ of the number $N_B$ of BS deployed at optimal locations $\mathbf{d}^*$ with optimal sectoring type $\mathcal{M}^*$ and power allocation $P_t^*$. Algorithm 1 starts with the calculation of optimal location $\mathbf{d}^*$ of BSs and sectoring $\mathcal{M}^*$ for a given value of $N_B$, and then determines the optimal power allocation $P_t^*$ at $\mathbf{d}^*$ in sectoring $\mathcal{M}^*$. Finally, it returns the operational cost $f(N_B)$ at $P_t^*$ for a given value of $N_B$. Using this $f(N_B)$,

---

**Algorithm 1** Calculating operational cost function $f(N_B)$ for a given $N_B$ and predefined set of system papameters

---

**Input:** Number of BSs $N_B \leq N_{B,max}$ and all other system parameters as defined in Section II

**Output:** Operational power cost

  1: Using Table I, find the optimal location $\mathbf{d}^*$ and sectoring $\mathcal{M}^*$ for a given value of $N_B$

  2: Set $j = 1$ for $\mathcal{M} = mk$ and $j = 0$ for $\mathcal{M} = mk + 1$

  3: **for** $i = j$ to $m$ **do**,

  4:      Calculate the minimum value of power allocation at $d_i^*$ in $i^{th}$ cell using (9)

  5: **end for**

  6: Using (10), find the optimal power allocation $P_t^*$ for all BSs by taking the maximum transmit power over $m$ values calculated in step 4

  7: Calculate the operational cost as $N_B(a_B P_t^* + b_B)$.

---

**Algorithm 2** Iterative scheme to obtain optimal operational cost

---

**Input:** Bounds $N_B^l$, $N_B^u$, and acceptable tolerance $\varsigma$

**Output:** Optimal operational cost along with joint optimal solution $(N_B^*, \mathbf{d}^*, \mathcal{M}^*, P_t^*)$

  1: Calculate $N_B^p = \lceil N_B^u - 0.618 \times (N_B^u - N_B^l) \rceil$

  2: Calculate $N_B^q = \lfloor N_B^l + 0.618 \times (N_B^u - N_B^l) \rfloor$

  3: Calculate $f(N_B^p)$ and $f(N_B^q)$ using Algorithm 1

  4: Set $\Delta_N = N_B^u - N_B^l$

  5: **while** $\Delta_N > \varsigma$ **do**

  6:      **if** $f(N_B^p) \leq f(N_B^q)$ **then**

  7:          Set $N_B^u = N_B^q$, $N_B^q = N_B^p$, and $N_B^p = \lceil N_B^u - 0.618 \times (N_B^u - N_B^l) \rceil$

  8:      **else**

  9:          Set $N_B^l = N_B^p$, $N_B^p = N_B^q$, and $N_B^q = \lfloor N_B^l + 0.618 \times (N_B^u - N_B^l) \rfloor$

  10:      **end if**

  11:      Calculate $f(N_B^p)$ and $f(N_B^q)$ using Algorithm 1

  12:      Set $\Delta_N = N_B^u - N_B^l$

  13: **end while**

  14: Calculate $N_B^* = \left\lceil \frac{N_B^u + N_B^l}{2} \right\rceil$

  15: Calculate optimal operational cost $f(N_B^*)$

  16: Using Table I, find the optimal location $\mathbf{d}^*$ and sectoring $\mathcal{M}^*$ by substituting $N_B = N_B^*$

  17: By substituting optimal deployment of BSs as obtained in steps 14 and 16 in equation (10), the optimal power allocation $P_t^*$ for all BSs is obtained

---

Algorithm 2 calculates the optimal number of BSs $N_B^*$ and corresponding minimized operational cost $f(N_B^*)$ by using golden section method that exploits the unimodality of operational cost $f(N_B)$ in $N_B$. The feasible search range of number of BSs $N_B$ lies between $N_B^l$ and $N_B^u$. We set the lower bound $N_B^l$ on $N_B$ as 1 and the upper bound $N_B^u$ as the maximum number $N_{B,\max}$ of BSs that are available for deployment based on the overall budget. This search space $N_B^u - N_B^l$ reduces by a fixed factor of $0.618$ after each iteration. The detailed steps followed in finding $N_B^*$ are mentioned in Algorithm 2. Since the objective is unimodal in $N_B$ and golden section algorithm is known to have fast convergence for unimodal functions [32], Algorithm 2 finds $N_B^*$ in very few iterations, which is also validated via numerical results in Section VII.

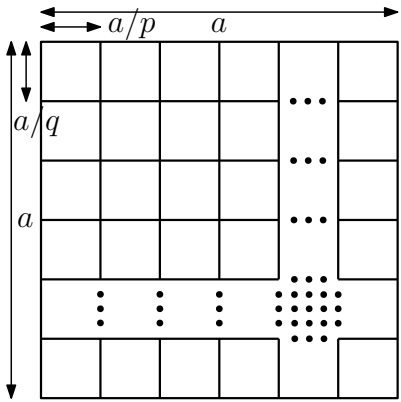

Fig. 5: Generation of cells along the length and width of a square field of side length $a$.

## VI. Operational Cost Minimization in a Square Region

As a square field is symmetric along its length and width, the cells are generated without any coverage hole by dividing the field along its length and width as shown in Fig. 5, where the length and width are divided into $p$ and $q$ equal segments, respectively, i.e., the number of cells $= pq = N_B$. The generated cells are square if $p = q$, otherwise they are rectangle in shape. It is evident that rectangular and square cells are equally divided by two orthogonal axis along the length and width with center as its origin, so the optimal location of the BSs are at the center of the cells for any value of $N_U$ (cf. Section IV).

We aim to obtain the condition on $p$ and $q$ for finding the minimum value of farthest point Euclidean distance from the center. If we relax the integer value of $p$ and $q$, and set $q = p - x$ ($x \geq 0$), then farthest point euclidean distance in a cell is $r_{u,c} = \frac{a}{2}\sqrt{\frac{1}{p^2} + \frac{1}{(p-x)^2}}$, where $a$ is side length of the square field. Minimum value of $r_{u,c}$ depends on the minimum value of $\mathcal{D} = \frac{1}{p^2} + \frac{1}{(p-x)^2}$. As $N_B = p(p-x) \Rightarrow p = \frac{x + \sqrt{x^2 + 4N_B}}{2}$, $\mathcal{D} = \frac{x^2 + 2N_B}{N_B^2}$, $\frac{\partial \mathcal{D}}{\partial x} = \frac{2x}{N_B^2}$, $\frac{\partial^2 \mathcal{D}}{\partial x^2} = \frac{2}{N_B^2} > 0$, i.e., $r_u$ is convex and achieves a minimum value at $x^* = 0$, i.e., at $p = q$. But for a given $N_B$, if $p = q$ is not possible for the integer value of $p$ and $q$, then $N_B = pq$ should be such that $x = |p - q|$ must be a minimum possible integer. Next, we discuss about the operational cost minimization in a square field under two different UEs density scenarios.

## A. When the Density of UEs is Asymptotically Very High

As discussed in Section IV, $f_{\text{far},c}(r_{\text{far},c}) = \delta(r_{\text{far},c} - r_{u,c})$ for $N_U \to \infty$, where $f_{\text{far},c}(r_{\text{far},c})$ is PDF of distance $r_{\text{far},c}$ of farthest UE in a cell of the square field. Here a cell is denoted by a suffix $c$, because all the cells are same. Also, we have dropped the location of the BS in the PDF expression as it is trivial that the BSs are lying at center of the cells. For coverage above 90% ($\epsilon \leq 0.1$), the tight bound of power allocation over the BSs is $P_t \geq \frac{T\sigma^2}{\epsilon} \int_0^{r_{u,c}} r_{\text{far},c}^\alpha f_{\text{far},c}(r_{\text{far},c}) \mathrm{d}r_{\text{far},c} = \frac{T\sigma^2}{\epsilon} r_{u,c}^\alpha$ as discoursed in Section V-A. For $N_B = pq$, $r_{u,c} = \frac{a}{2}\sqrt{\frac{1}{p^2} + \frac{1}{q^2}}$; therefore, optimal power allocation in each cell is $P_t^* = \frac{T\sigma^2}{\epsilon}\left(\frac{a}{2}\sqrt{\frac{1}{p^2} + \frac{1}{q^2}}\right)^\alpha$. Now problem (P0) can be written as:

$$\text{(P1): } \underset{N_B}{\text{minimize}} \quad N_B\left[c_B\left(\frac{a}{2}\sqrt{\frac{1}{p^2} + \frac{1}{q^2}}\right)^\alpha + b_B\right] \tag{11}$$

$$\text{subject to} \quad N_{B,\max} \geq N_B > 0, P_{t,\max} \geq P_t > 0$$

where $c_B = \frac{T\sigma^2}{\epsilon}$. The objective function of problem (P1) is unimodal and pseudoconvex with respect to $N_B$, which can be explained same as the discourse in Section V-B. One of the method for determining the minimized operational cost is by defining a function $f(N_B)$ like in Algorithm 1 in which first we find the optimal way of division of cells in the square field for a given value of $N_B$ for minimization of farthest point Euclidean distance, i.e., $N_B = pq$ such that $|p - q|$ is a minimum possible integer. Then we can find the $P_t^* = \frac{T\sigma^2}{\epsilon}\left(\frac{a}{2}\sqrt{\frac{1}{p^2} + \frac{1}{q^2}}\right)^\alpha$ and corresponding operational cost as a output for a given value of $N_B$. Finally using the Algorithm 2, we can find $N_B^*$ and optimal operational cost. The drawback of this approach is that we have to apply Algorithm 2 over whole range of $N_B \in [1, N_{B,\max}]$ In this case, we can make Algorithm 2 more efficient by reducing the range of $N_B$ by a very large factor for the iteration. First we take those $N_B$ values in which $N_B = p^2$ ($p = q$), then we find closed form solution of optimal number of BSs $N_{B,p^2}^*$. Lastly, we restrict the range of $N_B$ around $N_{B,p^2}^*$ for the iterative approach of determining $N_B^*$.

For $N_B = p^2$, the problem (P1) can be written as:

$$\text{(P2): } \underset{N_B}{\text{minimize}} \quad N_B\left[c_B\left(\frac{a}{\sqrt{2N_B}}\right)^\alpha + b_B\right] \tag{12}$$

$$\text{subject to} \quad N_{B,\max} \geq N_B > 0, P_{t,\max} \geq P_t > 0.$$

Double differentiation of the objective function of (P1) is $\frac{\alpha(\alpha-2)c_B a^\alpha}{N_B^{(\alpha/2+1)}2^{(\alpha/2+2)}} > 0$ for $\alpha > 2$. Therefore, the objective function is strictly convex and has global minimum value at $N_{B,p^2}^* = \frac{a^2}{2}\left\{\frac{(\alpha/2-1)c_B}{b_B}\right\}^{2/\alpha}$. As $N_{B,p^2}^*$ may be a fractional value, the restricted range of $N_B$ for iterative solution is $\left[\left\lfloor\sqrt{N_{B,p^2}^*}\right\rfloor^2, \left\lceil\sqrt{N_{B,p^2}^*}\right\rceil^2\right]$ for $P_{t,\max} < \frac{b_B C_B^{\alpha-1}}{\alpha/2-1}$ and $\left[\left\lfloor\sqrt{\frac{1}{2}\frac{(aC_B)^2}{(P_{t,\max})^{2/\alpha}}}\right\rfloor^2, \left\lceil\sqrt{\frac{1}{2}\frac{(aC_B)^2}{(P_{t,\max})^{2/\alpha}}}\right\rceil^2\right]$ for $P_{t,\max} > \frac{b_B C_B^{\alpha-1}}{\alpha/2-1}$ For $\alpha = 2$, the objective function is a affine transform of $N_B$ which gives $N_B^* = \max\left\{1, \left\lceil\frac{1}{2}\frac{(aC_B)^2}{P_{t,\max}}\right\rceil\right\}$.

### B. When the Density of UEs is Moderate

Here the PDF of distance of farthest UE in a cell, as obtained using (4), is given by:

$$f_{\text{far},c}(r_{\text{far},c}) = \sum_{z=0}^{N_U}\binom{N_U}{z}\left(\frac{1}{N_B}\right)^z\left(1-\frac{1}{N_B}\right)^{(N_U-z)} \tag{13}$$
$$\times z f_c(r_{\text{far},c}) \times \left[F_c(r_{\text{far},c})\right]^{(z-1)},$$

where $f_c(r_{\text{far},c})$ and $F_c(r_{\text{far},c})$ are the PDF and CDF respectively of distance $r_{\text{far},c}$ of an UE in a cell. Corresponding optimal power allocation in every cell is $P_t^* = \frac{T\sigma^2}{\epsilon}\int_0^{r_{u,c}} r_{\text{far},c}^\alpha f_{\text{far},c}(r_{\text{far},c})dr_{\text{far},c}$ for coverage above $90\%$ (cf. Section V-A). In this case the problem (P0) can be written as:

$$\text{(P3): } \underset{N_B}{\text{minimize}} \ \ N_B\left[c_B\int_0^{r_{u,c}} r_{\text{far},c}^\alpha f_{\text{far},c}(r_{\text{far},c})dr_{\text{far},c} + b_B\right] \tag{14}$$

$$\text{subject to} \ \ N_{B,\max} \geq N_B > 0, P_{t,\max} \geq P_t > 0,$$

Similar to discussion in Sections V-B and VI-A, the objective function of problem (P3) is unimodal and pseudoconvex with respect to $N_B$. Here also, first we take $p = q$, i.e., $r_{u,c} = \frac{a}{\sqrt{2N_B}}$, then define $f(N_B) = N_B\left[c_B\int_0^{\frac{a}{\sqrt{2N_B}}} r_{\text{far},c}^\alpha f_{\text{far},c}(r_{\text{far},c})dr_{\text{far},c} + b_B\right]$. Using Algorithm 2, we find the optimal number of BSs $N_{B,p^2}^*$. Now, we can restrict the range of $N_B$ to $\left[\left(\sqrt{N_{B,p^2}^*}-1\right)^2, \left(\sqrt{N_{B,p^2}^*}+1\right)^2\right]$. After redefining objective function of (P3) as $f(N_B)$, we can find the optimal number of BSs $N_B^*$ and minimized operational cost using Algorithm 2 in the restricted range of $N_B$. Next, we will obtain the numerical results of proposed analytical system for minimization of operational cost.

## VII. NUMERICAL RESULTS AND DISCUSSION

Now we conduct numerical investigation on the proposed optimization and solution methodology. The simulation parameters considered are: $R = 500$ m, $a = 500\sqrt{\pi}$ m, $N_U = 120$, $T = -10$ dB, $\alpha = 4$, $\sigma^2 = -70$ dBm, $a_B = 5.5$, $b_B = 32$ W, $P_{t,\max} = 5$ W, and $N_{B,\max} = 35$. In fixed allocation scheme for experiments over the circular field, $P_t = 4$ W, $N_B = 35$, sectoring $\mathcal{M} = k$ ($m = 1$), location $d_1 = 250$ m for $N_B \geq 2$; $d_1 = 0$ for $N_B = 1$.

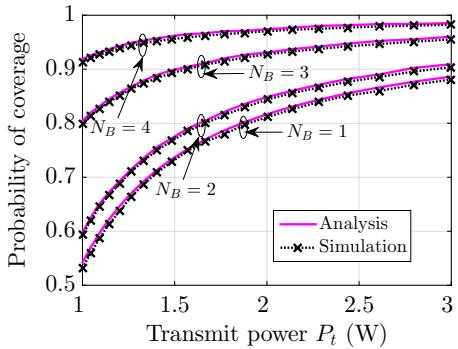

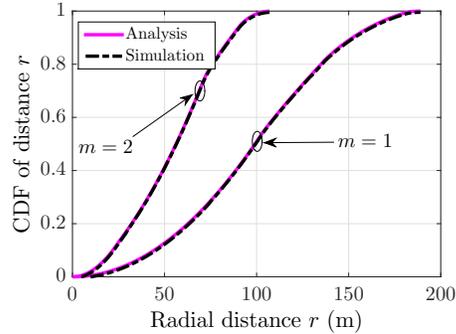

Fig. 6: Variation of average coverage probability with transmit power of BSs for different $N_B$.

Fig. 7: Variation in CDF of distance $r$ in $m^{th}$ cell of $\mathcal{M} = mk + 1$ ($k = 10$) for different values of $m$.

### A. Validation of Analysis

Firstly, we validate the average coverage probability expression given in (5). For validation, the simulation results are obtained by first examining $10^6$ random realizations of Rayleigh fading channel gain for the corresponding received SNR at the farthest UE in a cell to be greater than $-10$ dB for a given UE deployment. After that the average of this fraction, for which SNR $\geq -10$ dB, is taken over the $10^3$ random UE deployments. A closed match between analytical and simulation as observed in Fig. 6, validates the average coverage probability analysis as discussed in Section III-A. We also verified the quality of approximation (9) for the average coverage probability constraint $C1$ by noting that the corresponding root-mean-square error was less than $0.018$ for $N_B = 4$. As mentioned in Section V-A, this approximation error diminishes very rapidly with decreasing threshold $\epsilon$.

From Fig. 6, we also observe that there is not much improvement in the average coverage probability when two BSs are deployed instead of one BS. This is so because as center is optimal BS location in both cases, there is no reduction in the distance of the farthest point inside a cell from its BS on increasing $N_B$ from 1 to 2. Through a high improvement in average coverage probability, when $N_B$ is increased from 2 to 3, the improvement margin again decreases for $N_B = 4$. So, we note that when the cells are generated by dividing the field in angular direction only, then the reduction rate of farthest point distance from the BS decreases with $N_B$. So, for higher improvement in reduction of the farthest point distance, we move to higher sectoring types, where the cells are generated in both angular and radial directions (cf. Fig. 1).

Also, we have validated the proposed CDF in appendix A of distance $r$ of an UE from the BS in $m^{th}$ cell for sectoring $\mathcal{M} = mk + 1$ where $k = 10$. It can be observed in Fig. 7 that the CDF reaches to value 1 at a faster rate for higher $m$ because the farthest point Euclidean distance from the BSs in each cell is reducing rapidly with $m$.

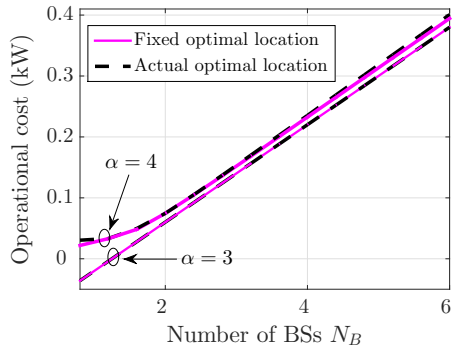

Fig. 8: Difference in the operational costs of the BSs deployed in two optimal scenarios.

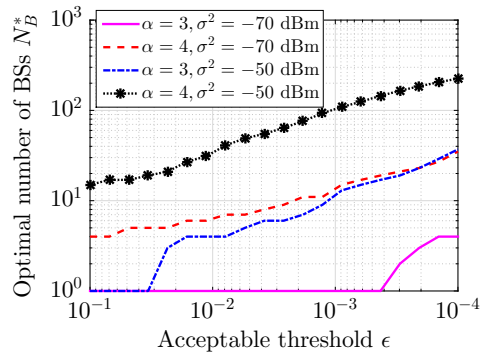

Fig. 9: Variation of optimal number of BSs with threshold $\epsilon$ for different values of $\alpha$ and $\sigma^2$.

Lastly, we investigate the quality of approximation for the optimal BS location based on the ideology of minimizing the Farthest point Euclidean distance in each cell. In this regards, in Fig. 8 we have plotted the operational cost performance for (a) BSs localization based on the minimization the Farthest point Euclidean distance in each cell (called fixed optimal location) and (b) BSs localization based on the optimal locations (called actual optimal location) as found by searching in the neighborhood of the ones that minimize the Farthest point Euclidean distance in each cell (cf. Section IV). As observed in Fig. 8, for both $\alpha = 3$ and $\alpha = 4$, the operational cost with fixed optimal location is in close match with the one for the optimal BS location. This validates our proposal and therefore, hereafter the deployment of BSs based on fixed optimal location has been taken in our experiments for simplicity.

### B. Impact of System Parameters on Operational Cost versus Coverage Demand Trade off

Now, we investigate the impact of channel conditions ($\alpha$ and $\sigma^2$) on optimal number of BSs $N_B^*$ as obtained using the proposed joint optimization. As shown in Fig. 9, generally the increase in coverage demand, as represented by decreasing $\epsilon$, results in a significant increase in $N_B^*$. However, for $\alpha = 3$ and $\sigma^2 = -70$ dBm that represents the most favorable channel conditions, $N_B^* = 1$ is sufficient for meeting high coverage quality demand with thresholds upto $\epsilon \geq 4.3 \times 10^{-4}$. For, $\alpha = 3$, an increase in $\sigma^2$ from $-70$ dBm to $-50$ dBm results in an average increase of about $5.53$ times in $N_B^*$. Similarly, when $\alpha$ increases from 3 to 4 for $\sigma^2 = -70$ dBm, $N_B^*$ on an average gets increased by $7.6$ times. Thus, the optimal BSs requirement $N_B^*$ not only depends on the coverage quality threshold $\epsilon$, but is also strongly affected by the channel conditions $\alpha$ and $\sigma^2$.

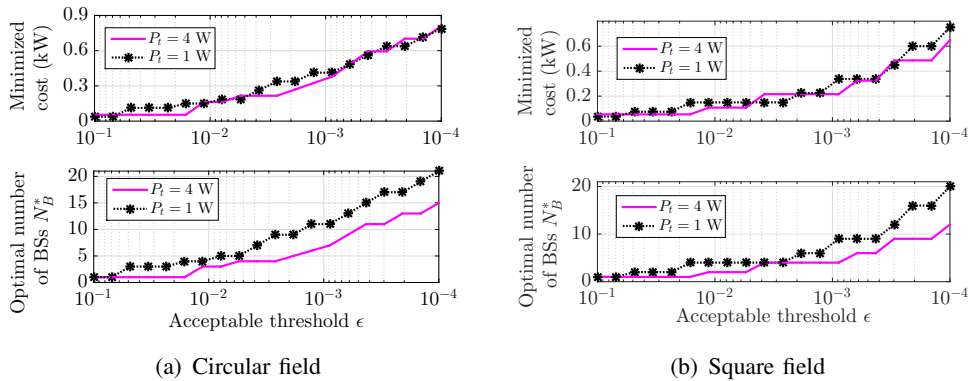

Fig. 10: Variation of minimized cost and optimal number of BSs with $\epsilon$ for different values of $P_t$.

In Fig. 10, we have plotted the tradeoff between the minimized operational cost and the underlying average coverage probability requirement $1 - \epsilon$ in a circular and square field for $\sigma^2 = -80$ dBm. Here, $N_B$ and BS locations are jointly optimized for a given $P_t$ value. We notice that the optimal number of BSs $N_B^*$ is lower for higher $P_t$ and vice versa. This results in almost the same cost for the two $P_t$ values, because, for lower $P_t$, higher number of BSs are deployed and for higher $P_t$, $N_B^*$ is relatively lower. As acceptable average coverage probability increases from $0.9$ to $0.9999$, the corresponding cost increasing from $40$ W to $800$ W and from $40$ W to $700$ W corroborates the utility of the proposed framework for $\epsilon \ll 1$ in the circular and square fields, respectively.

The plots of optimal number of BSs $N_B^*$ with transmitted power $P_t$ for satisfying the thresholds $\epsilon = 10^{-2}$ and $\epsilon = 10^{-3}$ have been shown in Fig. 11. At $P_t = 0.25$ W and $P_t = 5$ W, we require 18 and 9 more optimal number of BSs respectively for $\epsilon = 10^{-3}$ than $\epsilon = 10^{-2}$. Therefore, the requirement of more optimal number of BSs on average decreases with increment in $P_t$ and both the curves will asymptotically converge to $N_B^* = 1$ for very high value of $P_t$.

As shown in Fig. 12, the rate of increment in operational cost with number of UEs $N_U$ increases with increment in coverage demand. Also, we can observe that when the number of optimal number of BSs $N_B^*$ is constant with increment in $N_U$, then the optimal transmitted power

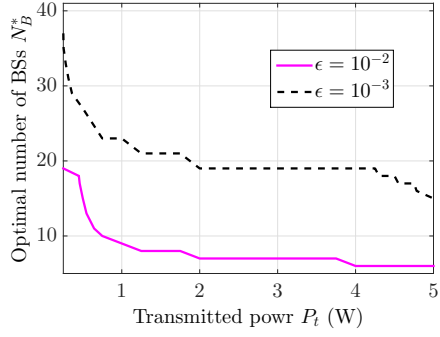

Fig. 11: Variation of optimal number of BSs $N_B^*$ with $P_t$ for different values of $\epsilon$.

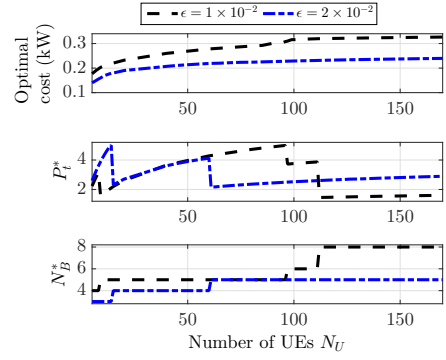

Fig. 12: Variation of optimal operational cost, $P_t^*$, and $N_B^*$ with number $N_U$ of UEs.

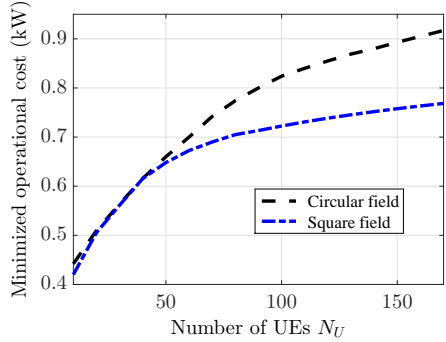

(a) Variation with number $N_U$ of UEs

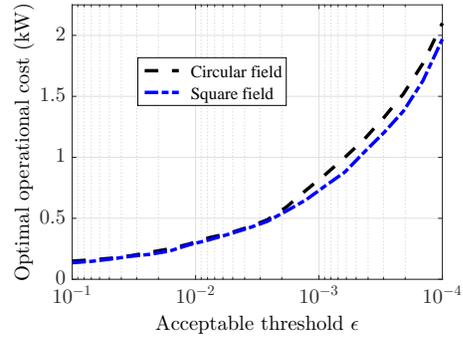

(b) Variation with acceptable threshold $\epsilon$

Fig. 13: Comparison of minimized operational cost in the circular and square fields.

$P_t^*$ increases upto $P_{t,\max} = 5$ W and suddenly drops, when $N_B^*$ increases. Therefore, there is no sudden change in the minimized operational cost due to trade-off nature between $N_B^*$ and $P_t^*$ as discussed in Fig. 10.

### C. Performance Comparison Results

In Fig. 13, we have compared the minimized operational cost in a circular and square field of same area. Initially, they follow each other with respect to $N_U$ and the difference gradually enhances after $N_U = 40$ for satisfying the threshold $\epsilon = 10^{-3}$ at $\sigma^2 = -70$ dBm as shown in Fig. 13(a). The minimized operational cost in circular field is $12$ W and $148.5$ W higher than in square field for $N_U = 50$ and $N_U = 170$, respectively. But with respect to threshold $\epsilon$, the minimized operational cost is almost same for $\epsilon \geq 0.002$ otherwise minor increment in the cost occurs in circular field for $N_U = 120$ as depicted in Fig. 13(b).

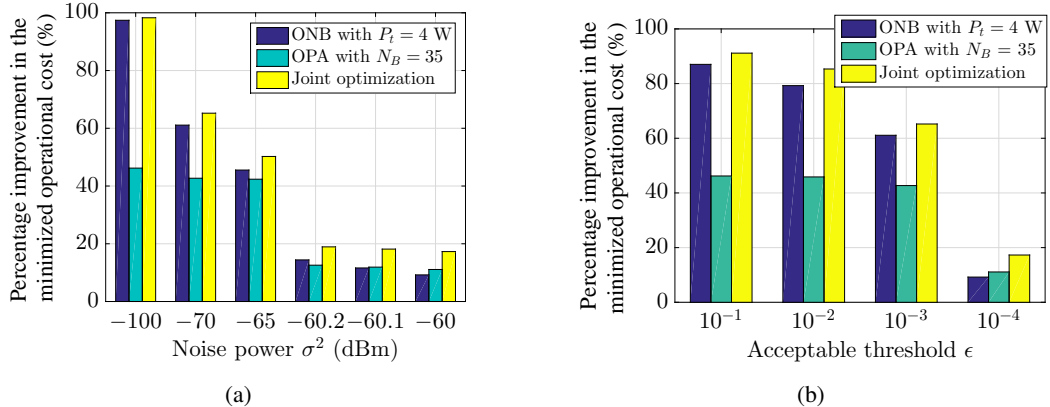

Fig. 14: Percentage improvement provided by the different optimization methods with varying (a) $\sigma^2$ and (b) $\epsilon$.

Finally, we conduct a comparison study in the circular field, where the relative performance of three optimization schemes, (i) ONB: optimal number of BSs for $P_t = 4$ W, (ii) OPA: optimal power allocation for $N_B = 35$, and (iii) proposed joint optimization, are compared against the fixed allocation scheme. The optimization with respect to the BS locations and sectoring type is considered for all three schemes under different noise power $\sigma^2$ and acceptable threshold $\epsilon$ settings. From Figs. 14(a) and 14(b), we note that ONB has better performance than OPA for $\sigma^2 \leq -60.1$ dBm and $\epsilon = \{10^{-1}, 10^{-2}, 10^{-3}\}$, respectively. However, for very high noise power $\sigma^2 \geq -60$ dBm or for very high coverage demand with $\epsilon = 10^{-4}$, a large number of BSs are needed to be deployed. This happens because as OPA is already having very high $N_B = 35$, which is near optimal for $\sigma^2 \geq -60$ dBm and $\epsilon = 10^{-4}$, the optimization with respect to $P_t$ helps OPA in achieving better performance than ONB for higher noise power scenarios or higher QoS applications (lower $\epsilon$). The best performance is undoubtedly achieved by the proposed joint optimization scheme, which yields an average reduction of about $65\%$ in the operational cost with varying QoS or coverage demands (represented by varying $\epsilon$) as compared to the fixed allocation scheme.

## VIII. CONCLUDING REMARKS

We have efficiently solved the non-convex combinatorial operational cost minimization problem by using a novel solution methodology that involves decoupling of the joint practical problem into three individual optimization problems. Firstly, insights on optimal BS location and sectoring type were provided. A tight approximation for transmit power allocation was presented for high coverage demands with $\epsilon \leq 0.1$. Lastly, the optimal number of BSs was found iteratively by exploiting the global-optimality in $N_B$. Later, we have extended the methodology in a square field for finding the minimized operational cost. Numerical results presented insights on the impact of various system parameters on the tradeoff between the optimized cost and coverage quality. It is observed that the proposed joint optimization framework, yielding a significant performance enhancement over the benchmark schemes, can help in the practical realization of green QoS-aware network operation. Also, a square field has better performance than a circular field in minimization of the operational cost.

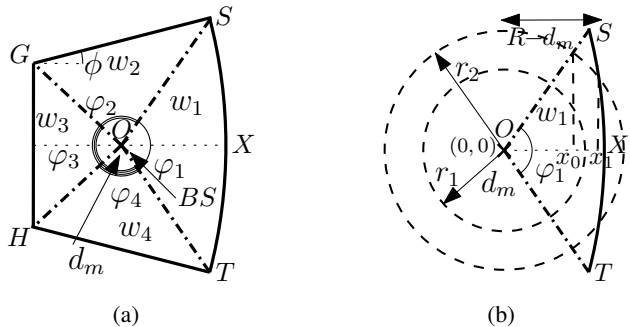

Fig. A.1: (a) Division of the $m^{th}$ cell in four disjoint sub-fields $\{w_j\}$; $j \in \{1,2,3,4\}$, (b) computation of distance distribution of an UE from location $d_m$ of the BS over the sub-field $w_1$.

## Appendix A

### Distance Distribution in $m^{th}$ Cell having Circular Arc Shaped Boundary

The possible shape of the cells generated by optimally deploying multiple BSs in a circular field are polygons except $m^{th}$ cell in which one of the boundary is an arc of the circular field as shown in Fig. 1. Calculation of distance distribution in polygon shaped fields has been discussed in [30], where all the boundaries are straight-line segment. But, no discussion has been done in fields associated with a circular arc shaped boundary. Here, we have extended the procedure in [30] for calculation of distance distribution in $m^{th}$ cell.

Similar to [30], we have divided the $m^{th}$ cell field $W_m$ in four disjoint sub-fields $\{w_j\}$; $j \in \{1, 2, 3, 4\}$ about the location $d_m$ of the BS associated with each boundaries as shown in Fig. A.1(a), i.e. $W_m = w_1 \cup w_2 \cup w_3 \cup w_4$. The distance distribution of an UE from the location $d_m$ i.e., the CDF $\frac{A_j(r)}{\mathbf{Ar}(W_m)}$ and PDF $\frac{\partial A_j/\partial r}{\mathbf{Ar}(W_m)}$; $j \in \{1, 2, 3, 4\}$ [33] in each sub-field is computed individually, where $\mathbf{Ar}(W_m)$ is the area of the field $W_m$ and $A_j(r) = \mathbf{Ar}(C(d_m, r) \cap w_j)$ is the area of intersection between the two fields: a bounded circle $C(d_m, r)$ with radius $r$ centered at $d_m$ and sub-field $w_j$. Finally, we determine the overall distance distribution over the $m^{th}$ cell field $W_m$ by summing all the distance distribution in the four sub-fields, which are given by:

$$F_m(r, \mathbf{d_m}) = \frac{1}{\mathbf{Ar}(W_m)} \sum_{j=1}^{4} A_j(r), \qquad f_m(r, \mathbf{d_m}) = \frac{1}{\mathbf{Ar}(W_m)} \sum_{j=1}^{4} \frac{\partial A_j(r)}{\partial r}, \qquad \text{(A.1)}$$

where $\mathbf{d_m} = \{d_{m-1}, d_m\}$ is the location of BSs in $(m-1)^{th}$ and $m^{th}$ cells, respectively, $F_m(r, \mathbf{d_m})$ and $f_m(r, \mathbf{d_m})$ are the CDF and PDF respectively of distance $r$ of an UE from the location $d_m$ of the BS. As the sub-fields $w_2$, $w_3$, and $w_4$ are associated with straight-line segment shaped boundaries, the computation of distance distribution is same as discussed in [30]. Here we discuss the computation of distance distribution in the sub-field $w_1$ associated with a circular arc shaped boundary in a especial scenario, when the BS lie at the symmetric axis of the cell.

For more discernment, we have explored the sub-field $w_1$ in Fig. A.1(b). As the BS always reside on the symmetric line of the cell, the sub-field $w_1$ is symmetric about the axis $OX$. The bounded circle $C(d_m, r)$ interact with the boundaries in two discrete ranges $r_1 \in [0, R - d_m)$ and $r_2 \in [R - d_m, \sqrt{R^2 + d_m^2 - 2Rd_m \cos \phi}]$. Using trigonometric relationship, the area of the intersection field $A_1(r) = \mathbf{Ar}(C(d_m, r) \cap w_1)$ can be computed as:

$$A_1(r) = \begin{cases} \frac{\varphi_1 r^2}{2}, & 0 \le r < R - d_m \\ \frac{\pi R^2}{2} + mx_0^2 + x_1\sqrt{r^2 - x_1^2} - x_0\sqrt{r^2 - x_0^2} \\ -(x_1 + d_m)\sqrt{R^2 - (x_1 + d_m)^2} + r^2 \arctan \frac{x_1}{r^2 - x_1^2} & R - d_m \le r \le \\ -r^2 \arctan \frac{x_0}{r^2 - x_0^2} - R^2 \arctan \frac{x_1 + d_m}{\sqrt{R^2 - (x_1 + d_m)^2}} & \sqrt{R^2 + d_m^2 - 2Rd_m \cos \phi} \end{cases} \qquad \text{(A.2)}$$

and corresponding first derivative with respect to $r$ can be expressed as:

$$\frac{\partial A_1(r)}{\partial r} = \begin{cases} \varphi_1 r, & 0 \le r < R - d_m \\ \left( \varphi_1 + 2 \arctan \frac{x_1}{\sqrt{r^2 - x_1^2}} - \pi \right) r & R - d_m \le r \le, \\ & \sqrt{R^2 + d_m^2 - 2Rd_m \cos \phi} \end{cases} \qquad \text{(A.3)}$$

where $m = \tan(\varphi_1/2)$, $\varphi_1 = 2\left\{ \pi - \arccos\left( \frac{d_m - R\cos\phi}{\sqrt{R^2 + d_m^2 - 2d_m R\cos\phi}} \right) \right\}$, $x_0$, $x_1$ are intersection points as shown in Fig. A.1(b) which are given as $x_0 = \frac{r}{\sqrt{m^2 + 1}}$; $x_1 = \frac{R^2 - r^2 - d_m^2}{2d_m}$, and $2\phi$ is the angle of the sector in which $m^{th}$ cell is one of the generated cell. Area of the field $W_m$ can be easily obtained as: $\mathbf{Ar}(W_m) = \phi R^2 - \frac{1}{4} \tan \phi (d_m + d_{m-1})^2$, where $d_{m-1}$ is the location of BS in adjacent $(m-1)^{th}$ cell. Now, after substituting (A.2) and (A.3) into (A.1), respectively, we get the CDF $F_m(r, \mathbf{d_m})$ and PDF $f_m(r, \mathbf{d_m})$ respectively of distance $r$ from the location $d_m$ of the BS in the field $W_m$. Note that in sectoring $k$, where $m = 1$, the field $W_m$ has three boundaries (two straight-line segment and one circular arc), which generates three disjoint sub-fields.